## Direct Observation of Wilsonian Electrons in Thunderstorms

**A.Chilingarian, T.Karapetyan, and B.Sargsyan**

**A. I. Alikhanyan National Lab (Yerevan Physics Institute)**
Alikhanyan Brothers 2, Yerevan, Armenia, AM0036

**Abstract**

We analyze electron-rich Thunderstorm Ground Enhancements (TGEs) detected at Aragats using the SEVAN-Light spectrometer, which can separate charged and neutral particles and reconstruct their energy spectra. The events exhibit large electron fluxes, with reconstructed energies extending to 45 MeV. Simultaneous meteorological observations reveal that all events occurred within a remarkably narrow atmospheric regime characterized by exceptionally low cloud-base heights, temperatures near 0°C, strong electric-field disturbances, and lightning. Comparisons with the complete Aragats TGE archive demonstrate that the direct detection of runaway electrons becomes possible only when the active acceleration region approaches within several tens of meters of detector altitude, allowing electrons to survive atmospheric attenuation. These results provide an experimental explanation for the century-long rarity of observations of Wilsonian electrons. The findings establish the atmospheric conditions required for direct observation of runaway electrons and contribute to the understanding of particle acceleration in natural electric fields.

**Plane Language summary**

One hundred years ago, Nobel Prize winner C. T. R. Wilson proposed that strong electric fields inside thunderstorms could accelerate electrons to very high energies. Although the gamma rays produced by these electrons are now routinely observed, direct detection of the electrons themselves has remained surprisingly rare. Using a detector capable of separating electrons from gamma rays, we measured high electron fluxes up to 45 MeV. Electron-rich events all occurred under nearly identical atmospheric conditions. The cloud base descended to only a few tens of meters above the detector, placing the instrument unusually close to the region where particles were accelerated. In these rare cases, the acceleration region was so close that many electrons survived and reached the detector. Our results explain why direct observations of Wilson's electrons have remained rare for a century. The electrons are not absent; rather, the atmospheric conditions required to observe them occur only rarely. The study identifies the conditions necessary for direct detection of runaway electrons and improves our understanding of how thunderstorms act as natural particle accelerators.

## 1. Introduction

Wilson's original proposal that thunderstorm electric fields can accelerate electrons remains a central idea in high-energy atmospheric physics. The relativistic runaway electron avalanches model (RREA; Gurevich et al., 1992; Babich et al., 2001; Alexeenko et al., 2002; Dwyer, 2003) explains how seed electrons can be accelerated and multiplied in sufficiently strong electric fields, producing both relativistic electrons and bremsstrahlung gamma rays. However, despite substantial theoretical progress and numerous observations of gamma-ray glows (Torii et al., 2011 ), terrestrial gamma-ray flashes (TGFs, Fishman et al., 1994), and thunderstorm ground enhancements (TGEs, Chilingarian et al., 2010,2011)), direct observations of the accelerated electrons themselves have remained surprisingly rare.

This discrepancy represents a long-standing observational paradox. If runaway electrons are responsible for the gamma-ray emissions routinely detected during thunderstorms, why are direct observations of these electrons so uncommon? Recent reviews and commentaries have emphasized that reliable detections of Wilsonian electrons remain exceptional even after a century of experimental efforts. The rarity of these observations has fueled ongoing debates about the interpretation of observed Aragats electron-rich events (Williams et al, 2023; Mailyan et al., 2026).

The historical background of surface RREA observations and the reasons for delayed recognition over many decades are detailed in Chilingarian (2014) and Chilingarian et al. (2025). Observing TGE electrons is much more difficult than observing the gamma radiation they emit. Electrons lose energy continuously in air and are rapidly attenuated, whereas gamma rays can propagate to much greater atmospheric depths.

This difference creates a fundamental observational asymmetry. Most ground-based TGEs are photon-dominated, even when the source mechanism is electron acceleration. A detector can be highly efficient at detecting charged particles, but after tens or hundreds of meters of propagation, the gamma-ray population dominates the observed count rate in particle detectors. For this reason, a claim of electron detection cannot be based solely on TGE intensity; it must also be supported by atmospheric geometry and, most importantly, by charged-neutral separation.

The central question addressed in this paper is which atmospheric and geometric conditions allow TGE electrons to survive the path from the acceleration region to the particle detectors on Aragats. We demonstrate that Aragats observations provide an experimental explanation for the century-long rarity of Wilsonian electron detections. The principal limitation is not the absence of runaway electrons in thunderstorms, but rather the rarity of atmospheric conditions that permit their survival to detector altitude. The results identify a distinct near-ground RREA regime and establish the physical conditions required for direct observation of atmospheric runaway electrons.

## 2. Method

We present six electron-rich TGEs detected by the SEVAN-Light spectrometer at Aragats during 2024–2026. By combining electron flux measurements, reconstructed electron energy spectra, cloud-base estimates, lightning observations, and near-surface electric-field measurements, we identify the atmospheric conditions that enable direct observation of relativistic electrons at ground level. The SEVAN Light (Space Environment Viewing and Analysis Network, Chilingarian et al., 2009; 2024a) is a compact, energy-deposit-resolving particle spectrometer that can separate neutral components (neutrons, γ-rays) from charged components (muons, electrons) of GLEs and TGEs (Sargsyan and Chilingarian, 2026a; Chilingarian et al., 2026b). SEVAN Light detectors installed at Aragats (3200 m) include a 0.25 $m^2$ spectrometric scintillator with a 20 cm thickness and a 1 cm thick, 1 $m^2$ veto scintillator. The detector incorporates high-dynamic-range logarithmic ADCs (LADCs) to measure deposits of 5–300 MeV and coincidence logic to classify events as charged or neutral particles. The vertical coincidence window is 1 μs, effectively separating accidental coincidences. SEVAN detectors are among the few instruments capable of simultaneously measuring count rates and energy spectra of both charged and neutral cosmic ray species. The electron energy is reliably recovered for electrons > 10 MeV. The samples selected by the "01" coincidence, which generate a signal in the lower scintillator but not in the upper, will be enriched in neutral particles. The samples selected by the “11” coincidence will be enriched in charged particles. The “11” coincidence channel is predominantly governed by TGE electrons, with a minor contribution from gamma rays. Electron registration efficiency increases from 60% to 99% as energy increases from 10 to 50 MeV. Correspondingly, gamma-ray contamination increases from 0.8% to 1.8%. Gamma rays of the same energy are registered by the “01” coincidence with an almost constant efficiency of ≈ 27%, and electron contamination is less than 0.1%.

For each registered particle burst, the following were stored: time series of particle count rates (electrons and gamma rays), percent enhancement, significance in σ units, cloud-base distance, distance to the nearest lightning flash, meteorological conditions, and near-surface electric field (NSEF). The baseline for the significance calculation was determined from quiet data preceding every single event. Cloud-base distance was estimated from the temperature and dew point using the standard approximation (SPREAD, 2026):

$H_{cloud} = 122 \times (T - T_d)$, with $H_{cloud}$ in meters and $T$ and $T_d$ in degrees Celsius. (1)

At Aragats, at 3200 m altitude during transition-season storms, surface air temperatures near 0 °C indicate that the station is close to the melting/mixed-phase layer and that the cloud base distance can descend to the station level. In the present analysis, the evidence for low cloud base and near-zero together indicates a low, mixed-phase cloud environment favorable for strong charge separation (Williams et al., 2023). The NSEF and lightning distance were taken from Boltek’s EFM-100 sensor. The archive comparison uses the full 2012-2025 TGE list. For archive-level means, a 15% two-sided trim was applied to temperature, distance to cloud base, and distance to lightning flash. This removes the tails of the broad distribution of the archive and yields robust estimates for the central TGE population. For the six electron-rich TGEs, the simple mean and RMSD are reported without trimming because the sample is small and the goal is to show the observed spread of the selected class rather than suppress it.

TGE events were identified directly from the continuous Aragats archive of synchronized 1-minute time series. A TGE candidate was defined as a sustained positive enhancement of the particle count rate above the running background, reaching several standard deviations above the local mean. Each candidate was then checked across the three independent scintillator channels to ensure a coherent detector response and to reject fluctuations in a single channel. Finally, the interval was confirmed as a TGE only if the synchronized near-surface electric field was disturbed, with |NSEF| > 5 kV $m^{-1}$. For each accepted event, the start time, duration, peak enhancement, detector significance, temperature, dew point, cloud-base distance, NSEF at maximum minimum and maximum NSEF, nearest lightning distance, and lightning-termination flag were extracted to form the event catalog (Chilingarian et al., 2024c).

## 3. Electron-rich TGEs

### 3.1 Overview of the six electron-rich TGEs

Figure 1 displays the TGEs in 2 panels: the percent enhancement of the count rate and the NSEF. This layout facilitates straightforward comparison of event morphology and atmospheric electric-field context. The individual event plots indicate that particle enhancements occur during periods of intense electric-field activity and are often abruptly ended by a lightning flash. Because direct observations of Wilsonian electrons remain rare, we briefly describe each of the six events in additional detail and present them in Table 1.

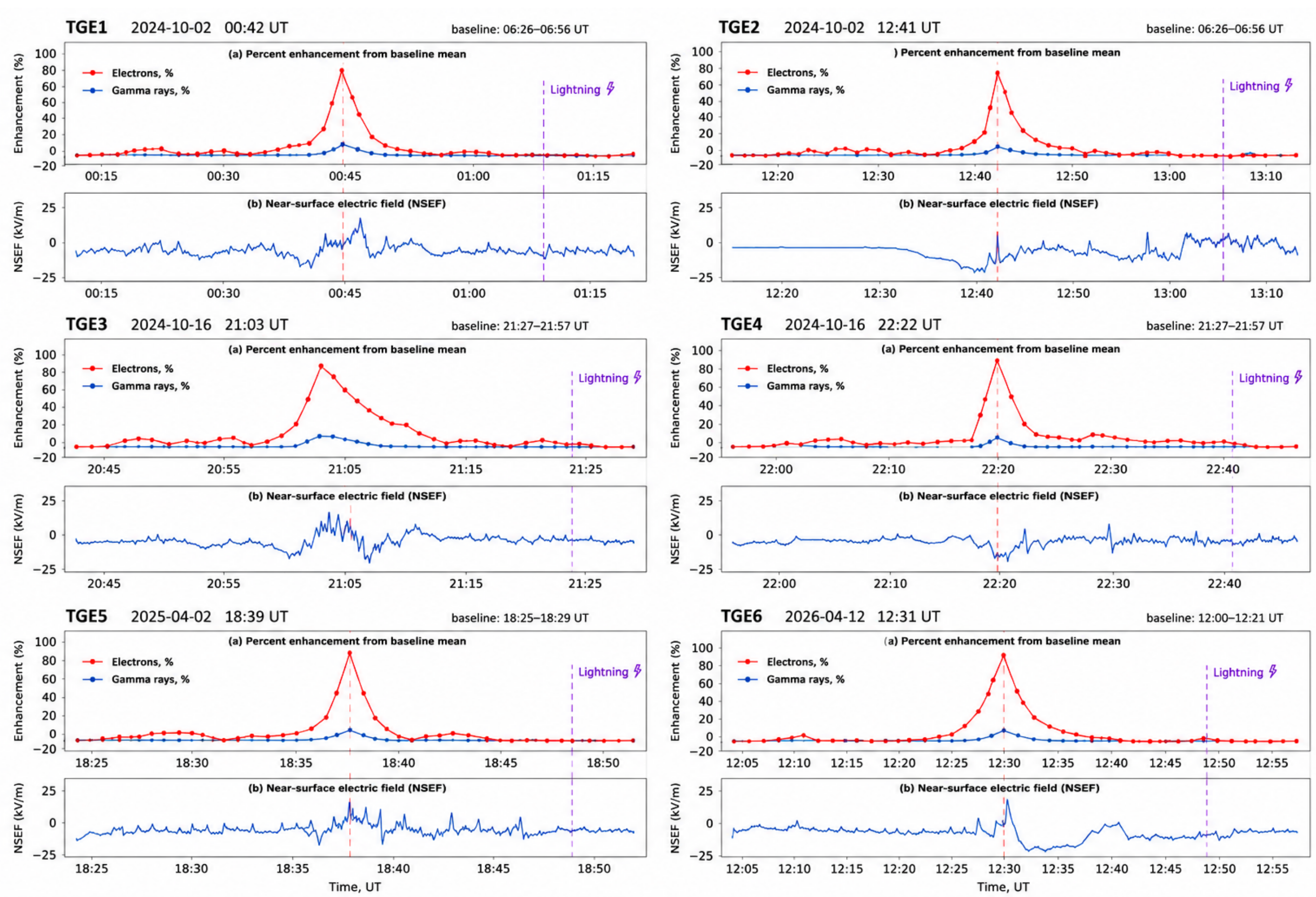

**Figure 1. Electron-rich TGEs: each event is shown with panels a-b: percent enhancement of count rate, and NSEF.**

TGE1, 2 October 2024 at 00:42 UT, is one of the two very strong gamma events. The gamma enhancement reaches 118.8% with 64.8 σ significance, while the electron channel increases by 22.3% with 19.0 σ significance. The cloud base is the highest among the selected electron events, at 87.5 m, but it is still extremely low compared with ordinary TGEs. The event is not classified as abruptly terminated by lightning, suggesting that the electric-field structure survived long enough for a short but intense enhancement. The NSEF is negative, -8.7 kV/m, consistent with a strong local field during the event.

TGE2, 2 October 2024, the same day at 12:42 UT, is the strongest event in the selected set by gamma enhancement: 134.7% and 73.5 σ. The electron enhancement is also the largest, 24.4% and 20.7 σ. It is lightning-terminated, has a cloud base of 50 m, and occurs during strong negative NSEF (-22.4 kV/m). This event is the clearest example of the combined condition identified in the paper: very low cloud base, strong NSEF, nearby lightning, and statistically strong electron response.

TGE3, 16 October 2024 at 21:03 UT, shows a more moderate gamma enhancement of 21.2% but still exhibits a strong, nearly equal electron signal of 19.3% and 14.9 σ. This demonstrates that the electron classification is not simply a consequence of extreme gamma brightness. The tabulated nearest flash distance is 14 km, making this event the main outlier in the lightning-distance column. However, the event still occurs at a very low cloud base (37.5 m) and positive NSEF (+11.1 kV/m), showing that both NSEF polarities can accompany electron TGEs.

TGE4, 16 October 2024 at 22:22 UT, is the second peak during the same storm period. It shows a gamma enhancement of 23.4% and an electron enhancement of 22.1%, both statistically very significant. The cloud base is only 25 m, the lowest in the sample, and the nearest flash is 3 km away. This event is lightning-terminated and therefore provides strong evidence that the particle flux is controlled by the persistence of the thundercloud electric field.

TGE5, 2 April 2025 at 18:39 UT, shows a gamma enhancement of 62.2% at 29.7 σ. The electron percent enhancement is smaller, 10.9%, with significance of 18.2 σ. The event occurs at cloud base 37.5 m and nearest flash distance 3 km, again placing it in the low-cloud, near-discharge regime.

TGE6, 12 April 2026 at 12:31 UT, is the coldest event in the sample (T = -2.9 C). It is not lightning-terminated, yet it still shows significant gamma enhancement (21.3%, 12.6 σ) and electron enhancement (12.4%, 10.8 σ). Its cloud-base distance is 37.5 m, and the nearest flash distance is 5 km. This event slightly broadens the temperature range but remains close to the mixed-phase regime, confirming that electron TGEs do not require the largest gamma intensities.

3.2 Event parameters

Table 1 summarizes the key particle and meteorological parameters of electron-rich TGEs. The gamma enhancement ranges from 21% to 135%, with a mean of 63.6%. The mean gamma significance is 35.8σ. The electron enhancement is smaller, at 18.6%. The electron significance is also highly consistent, at 16.8σ. This uniformity shows that the electron component is a statistically stable feature across the selected events. The event durations are short, 2-6 min, with a mean of 4.3 +/- 1.4 min. Four of the six events are classified as lightning-terminated. This supports the interpretation that these TGEs occur during a pre-discharge state of the thundercloud electric field. Lightning can rapidly neutralize or reorganize the field, terminating the avalanche process. The two non-terminated cases are still associated with nearby lightning but do not show the same immediate drop-in count rate.

The most compact meteorological variable is cloud-base distance. All six events have cloud-base distances between 25 and 87.5 m, with a mean of 44.7 ± 19.6 m. The lower boundary of the active thundercloud structure is very close to the detector. Such short distances are precisely the condition needed for relativistic electrons to reach the ground before being absorbed. The temperature distribution is very compact, 0.0 ± 1.3 °C, placing the events in or near the mixed-phase region where charge separation is efficient. NSEF has both signs and a broad RMSD; therefore, the sign of NSEF alone is not a discriminator. The field measurement is local and near-surface, whereas the acceleration field may be above the detector and embedded in a much more extended and complex charge structure.

**Table 1. Event-level parameters for the six electron-rich TGEs. The last row gives the simple mean and RMSD for the six events. The gamma and electron columns list percent enhancement and baseline σ significance. The table shows that the electron significance is consistently high, while cloud-base distance and temperature fall within a compact range favorable for short electron propagation distances.**

| Event | Peak UT | Duration min | gamma %/σ | e %/σ | Flash term. | NSEF kV/m | Flash km |
|---|---|---|---|---|---|---|---|
| TGE1 2024-10-02 | 00:42 | 2.0 | 118.8 / 64.8 | 22.3 / 19.0 | No | -8.7 | 3.0 |
| TGE2 2024-10-02 | 12:42 | 4.0 | 134.7 / 73.5 | 24.4 / 20.7 | Yes | -22.4 | 5.0 |
| TGE3 2024-10-16 | 21:03 | 6.0 | 21.2 / 16.2 | 19.3 / 14.9 | Yes | 11.1 | 14.0 |
| TGE4 2024-10-16 | 22:22 | 6.0 | 23.4 / 17.9 | 22.1 / 17.1 | Yes | -14.4 | 3.0 |
| TGE5 2025-04-02 | 18:39 | 4.0 | 62.2 / 29.7 | 10.9 / 18.2 | Yes | 1.2 | 3.0 |
| TGE6 2026-04-12 | 12:31 | 4.0 | 21.3 / 12.6 | 12.4 / 10.8 | No | -12.2 | 5.0 |
| Mean/RMSD | - | 4.3 +/- 1.4 | 63.6/35.8 +/- 47.1/24.3 | 18.6/16.8 +/- 5.1/3.2 | 4/6 Yes | - | 5.5 +/- 3.9 |

3.3 Recovered Energy Spectra

In Figure 2, we present the differential energy spectra of electrons measured above the roof of the SKL experimental hall, where the SEVAN Light spectrometer is positioned. The detector response was unfolded using a GEANT4-based response matrix, following the inverse-problem method described in Chilingarian et al. (2024b). The spectrum is well described by a simple exponential function of energy and aligns with the form given in Eq. 17 in Dwyer & Babich (2011). The maximum electron energy reaches 45 MeV.

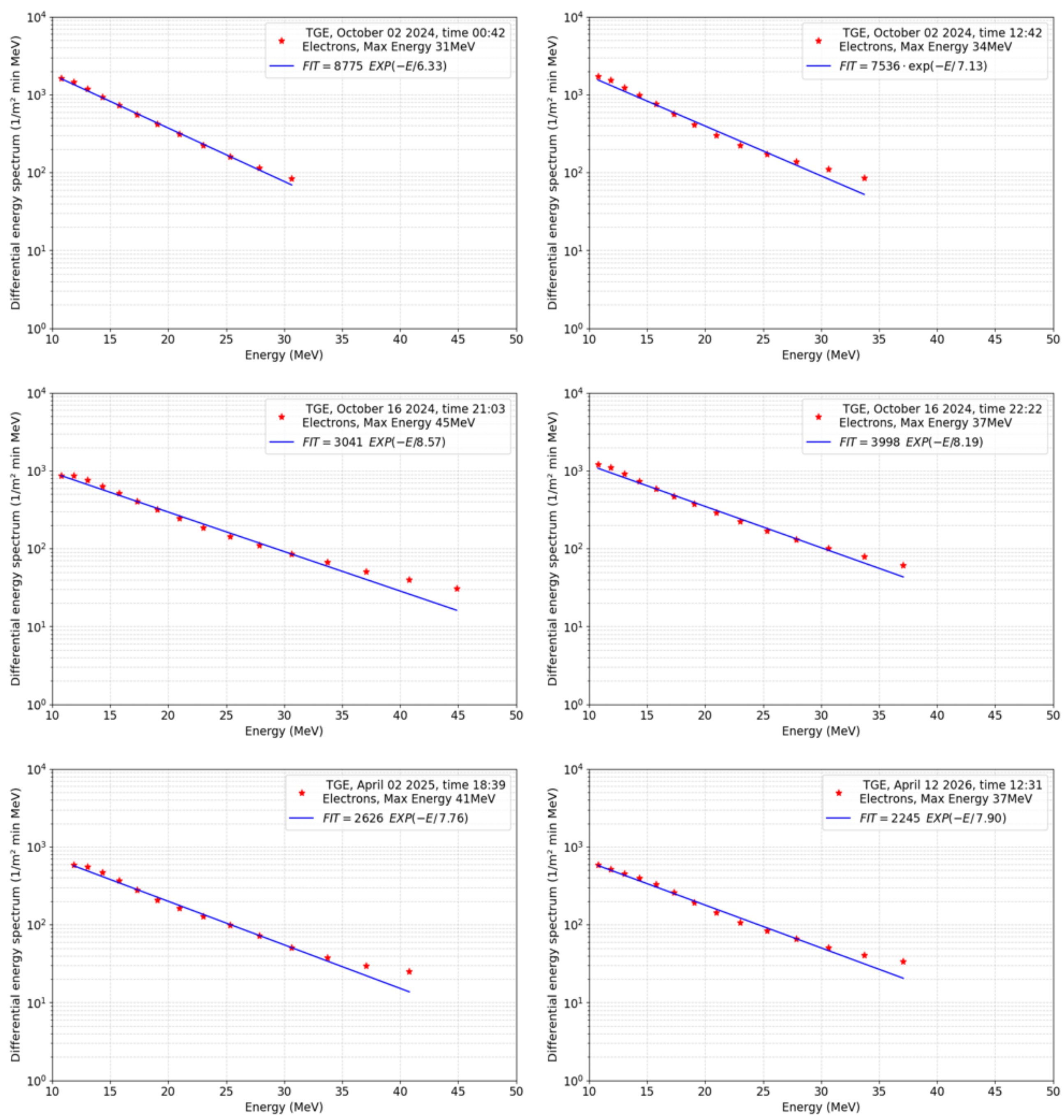

**Figure 2. Differential energy spectra of TGE electrons above the roof of the SKL experimental hall (SEVAN light spectrometer).**

The spectra shown in Figure 2 and Table 2 demonstrate that all six TGEs exhibit a remarkably stable exponential shape despite differences in intensity and meteorological conditions. The fitted characteristic energies $E_0$ vary only from 6.33 to 8.57 MeV, with a mean value of 7.65 ± 0.73 MeV. This stability suggests that the observed particles were generated by the same physical acceleration mechanism operating under slightly different atmospheric conditions. The maximum reconstructed electron energies reach 37–45 MeV in four of the six events. The highest reconstructed energies substantially exceed the energy range expected for ordinary environmental radioactivity or secondary detector effects.

**Table 2. Energy Spectra fit parameters with spread-derived cloud-base estimates**

| Event peak UT | Max Energy (MeV) | A parameter | $E_0$ parameter | T (°C) | $T_d$ (°C) | Spread $T-T_d$ (°C) | $H_{cloud}$ (m) |
|---|---|---|---|---|---|---|---|
| 2024-10-02 00:42 | 31.0 | 8775.0 | 6.33 | 0.8 | 0.1 | 0.70 | 85.4 |
| 2024-10-02 12:42 | 34.0 | 7536.0 | 7.13 | 0.4 | 0.0 | 0.40 | 48.8 |
| 2024-10-16 21:03 | 45.0 | 3041.0 | 8.57 | 0.9 | 0.6 | 0.30 | 36.6 |
| 2024-10-16 22:22 | 37.0 | 3998.0 | 8.19 | 0.8 | 0.6 | 0.20 | 24.4 |
| 2025-04-02 18:39 | 41.0 | 2626.0 | 7.76 | -0.1 | -0.4 | 0.30 | 36.6 |
| 2026-04-12 12:31 | 37.0 | 2245.0 | 7.90 | -2.9 | -3.2 | 0.30 | 36.6 |
| Mean ± RMSD | 37.5 ± 4.5 | 4703.5 ± 2524.0 | 7.65 ± 0.73 | -0.0 ± 1.3 | -0.4 ± 1.3 | 0.37 ± 0.16 | 44.7 ± 19.6 |

The spectral slopes correlate with the meteorological conditions summarized in Table 2. All six events occurred under nearly saturated atmospheric conditions, with very small temperature–dew-point spreads ($T-T_d$) of 0.2–0.7 °C. The estimated cloud-base heights ranged from only 25 to 85 m above detector level, with a mean of 44.7 ± 19.6 m. Under such conditions, relativistic electrons generated within the avalanche region can propagate directly to the detector without losing most of their energy.

The low cloud-base estimates also explain why the electron component was detectable simultaneously by several independent detector systems. In ordinary atmospheric cascades developing kilometers above ground, electrons are strongly attenuated, and the gamma component dominates at the detector level. In contrast, the present events developed mostly several tens of meters above the station. Therefore, both charged and neutral components of the avalanche could survive to detector altitude. An additional important result is the relative stability of the spectral form despite large differences in normalization parameter A. The parameter A varies from 2245 to 8775, reflecting large event-to-event differences in avalanche intensity, while $E_0$ remains comparatively stable. This indicates that the principal variation among events is the total multiplication of the avalanche rather than a fundamental change in the acceleration mechanism itself. Taken together, the spectral reconstruction, detector-response simulations, and meteorological analysis provide strong experimental evidence that the observed TGEs contained substantial relativistic electron fluxes that reached the Aragats detector level.

3.4 Correlations between meteorological parameters

Figure 3 shows the distribution of archives across three three-dimensional projections. Blue points represent TGEs from 2012-2025 with available temperature, cloud base, and lightning distance information. Red points mark the six electron-rich TGEs. The figure uses a visual overlay to keep the red markers visible without altering their physical coordinates. The purpose of the 3D figure is illustrative: it shows that the electron-rich TGEs occupy a compact cluster within the broader overall distribution.

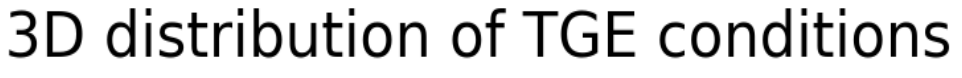

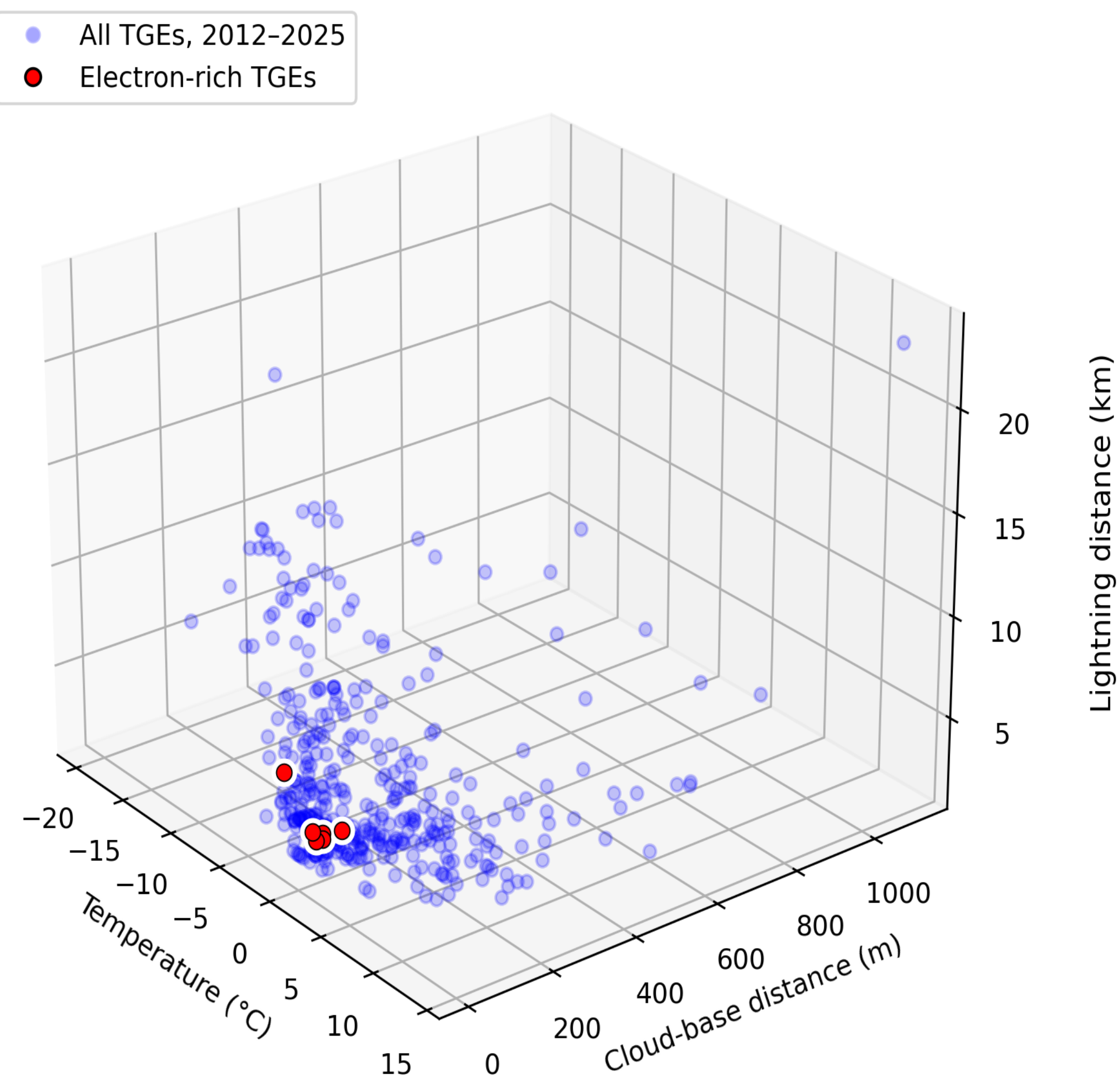

**Figure 3. Three-dimensional visualization of temperature, cloud-base distance, and lightning distance. Blue points are archive TGEs, and red points are the six electron-rich events. The red points are visually overlaid to remain visible inside the dense archive cloud. The figure confirms the compact, low-cloud-base region occupied by the selected events.**

## 4. Discussion and Conclusions

The six events observed from 2024 to 2026 provide strong experimental evidence for the direct detection of Wilsonian electrons at Aragats and identify the atmospheric conditions most favorable for their direct observation. More generally, the results demonstrate that electron-rich TGEs constitute a distinct near-ground RREA regime that occurs during low-base, mixed-phase, electrically active storms in the pre-discharge phase.

The six electron-rich TGEs form a remarkably compact population. Their mean cloud-base distance is only 44.7 ± 19.6 m, approximately a factor of 2.5 smaller than the corresponding value obtained from the central population of the 2012–2025 Aragats TGE archive. Temperatures cluster near 0°C, indicating that the events occur within the mixed-phase region of thunderclouds where charge separation processes are particularly efficient. Four of the six events were terminated by nearby lightning, demonstrating that the particle enhancements are closely connected with the evolution of the thundercloud electric field.

The reconstructed electron spectra provide independent evidence that relativistic electrons reached detector altitude. Maximum recovered energies extend to 45 MeV, while all six spectra exhibit nearly identical exponential shapes with characteristic energies between 6 and 9 MeV. The stability of the spectral form indicates that the same acceleration mechanism operated in all events and that the principal difference among events is avalanche intensity.

These observations resolve an important observational paradox in atmospheric high-energy physics. Since Wilson's original prediction, direct detections of runaway electrons have remained rare despite the widespread observation of thunderstorm gamma rays. The present results show that this rarity arises naturally from atmospheric geometry. Under ordinary thunderstorm conditions, electrons are strongly attenuated before reaching the ground, whereas gamma rays survive and dominate detector responses. Only when the acceleration region approaches within several tens of meters of the detector altitude can a substantial fraction of the runaway-electron population survive atmospheric propagation. Four of the six events terminate abruptly with lightning. When lightning neutralizes or reorganizes the field, the avalanche process rapidly ceases. The archive comparison supports the main physical conclusion of this study: detection of runaway electrons at ground level is controlled primarily by propagation distance and cloud geometry, rather than simply by the brightness of the TGE.

Direct observation of runaway electrons provides an experimental anchor for models of relativistic runaway electron avalanches, thunderstorm gamma-ray glows, terrestrial gamma-ray flashes, and lightning initiation. By identifying the atmospheric conditions under which runaway electrons are detected, the present work establishes evidence of an atmospheric electron accelerator operating in vast volumes of the thunderous atmosphere, with an impact on climate and global change yet to be estimated.

**Conflict of Interest**

The authors declare no conflicts of interest relevant to this study.

**Data Availability Statement**

The online data on continuous monitoring of meteorological parameters, electric field measurements, and particle fluxes at Aragats are available in graphical and numerical formats on the ADEI platform of CRD YerPhI: http://crd.yerphi.am/adei/